\documentclass[showpacs,preprintnumbers,amsmath,amssymb]{revtex4}
\usepackage{dcolumn}% Align table columns on decimal point %\usepackage{bm}% bold math
\begin{document}
\preprint{APS/123-QED}
\title{Superconductivity of the two-component
non-stoichiometric compounds with incommensurate sublattices.}
\author{V. N. Bogomolov}
\affiliation{A. F. Ioffe Physical \& Technical Institute,\\
Russian Academy of Science,\\
194021 St. Petersburg, Russia}
\email{V.Bogomolov@mail.ioffe.ru}
\date{\today}

\begin{abstract}
There exists a class of non-stoichiometric materials
(berthollides) that can be considered as constituted by two
sublattices, which have specific physicochemical properties. These
properties can be essentially modified by even rather weak
interaction between these components. One of them can be regarded
as a rigid matrix, while another one as a filling in the form of
isolated atoms, molecules or clusters. Structures containing voids
of the diameter up to $D \sim (1 - 2$)\,nm in diameter in the stoichiometric
sublattice belong to this class of compounds. These voids are
filled by the second component (of diameter $d_{0})$, which can be
compressed or stretched because of the sublattice parameters
misfit. A stretched matter $(D - d_{0} = h > 0)$ can exist in a unique
intermediate state between the metal and the dielectric; this
state cannot be implemented by another way. The period doubling occurs and
a weak modulation of the metal lattice constant leads to forming not only
the energy gap, but the bound electronic states of the molecular type
with two paired electrons as well.  Validity of this model
with the Peierls-type lattice instability for explanation of the
well known experimental data on superconducting transition
temperature $(T_{c})$ in such systems (fullerides, perovskite-type
compounds like Na-WO$_{3},$ high temperature superconductors) is
considered in this work. The transition temperature $T_{c}$ of
fullerides is proportional to $h/D$; for the tungsten-bronzes with Na,
Rb, or Cs, $T_{c} > 0$  for  $h > 0$, and  $T_{c} \sim 0$ for $h < 0$. 
\end{abstract}
\maketitle
\bigskip
      There exists a class of non-stoichiometric materials (berthollides) that can be considered as constituted by two sublattices, which have specific physicochemical properties \cite{bib1}. Such nanocomposites, nanostructures, cluster lattices can be found among natural non-stoichiometric compounds and can be fabricated by filling the structure voids of the stoichiometric matrices by various matters. Zeolites, clathrates, and perovskites on the one hand and fullerides, modified zeolites, intercalated compounds etc on the other hand, constitute this class of materials. Fullerides and some zeolites have slightly analogous structures, M$_{n}$C$_{60}$ and M$_{n}$[\rm{(Al, Si)}$_{24}$O$_{36}$], for instance, where M stands for atoms, molecules, or clusters of the second component. Lattices of spheres of diameter about 1\,nm composed from the carbon atoms or (Al,Si)O$_{4}$ tetrahedrons are present in both cases. Octahedral voids of diameter about 0.4\,nm between the spheres can !
be filled by the second component, elements of which M$_{n}$ are well separated one from another. Only one atom M finds room in MeO$_{6}$ octahedron tiny voids of perovskites. The second component atoms are located nearer with each other and can be considered as a weakly interacting sublattice ("subcondensate"). Specific properties of such systems stem both from the scale and geometric factors. Fullerides, for instance, not only have a large variety of normal properties, but also can be superconductors with $T_{c}$ up to 40K. Many of their properties are discussed in \cite{bib2}.

Zeolite-based nanocomposites and nanostructures also demonstrate a
wide spectrum of properties, which are not inherent to the
isolated second component matter. The nonlinear conductance, para-
and ferromagnetism, dielectric and optical properties, forming of
new molecular structure etc. are typical examples of their
features \cite{bib3}.

    A variety of properties of perovskites are well known and are related to some positional freedom of the second component atoms and their mobility within the matrix voids. Perhaps, just this is a reason of appearance of superconductivity with $T_{c} = 91$K in the
tungsten-bronze Na$_{x}$WO$_{3}$ \cite{bib4}.

   Sublattices of the components can be found obviously in a stressed state in these non-stoichiometric compounds due to possible misfit of their lattice constants. One of them will be compressed, and stretched another. The latter case is particularly important since an intermediate between the metal and the insulator state can appear that is of interest in the context of the high temperature superconductivity problem \cite{bib5}.
\paragraph*{\rm{I.}}
\label{pI}   Being stretched, a lattice formed from divalent metal atoms can transit into the dielectric state due to ceasing of the electronic bands overlapping and opening of the energy gap.
\subparagraph{1.}
\label{pI.1}
    A MeB$_{6}$ -kind compound is a
typical example. If Me stands for a divalent atom, the MeB$_{6}$
compound will be a dielectric, while it will be a metal in the
case of a monovalent metal (at the room temperature) \cite{bib6}.
Metal atom sublattice not just doping ions is introduced into
voids between B$_{6}$ octahedrons. Stretching of the matrix, when
the atomic diameter $d_{0}$ exceeds the MeB$_{6}$ lattice constant
$D$, indicates this. (See also point II 3).  The $D/d_{0}$ ratio
takes on the values for Me = Ce, La, Ca, Sr, Ba: Ce (0.414/0.326);
La (0.415/0.374); Ca (0.415/0.395); Sr (0.419/0.430); Ba
(0.427/0.434). The Ca sublattice is stretched and CaB$_{6}$ is a
dielectric. All of these compounds have close melting temperatures
$T \sim 2200$K, which are determined by the matrix B$_{6}$ \cite{bib7}.

Interatomic distances in the most of typical ionic compounds are
assumed to be equal to the sum of ionic radii, which can be
determined in a variety of ways (according to Pauling,
Goldschmidt, etc.). However, even in such a compound as NaCl, the
sum of ionic radii $0.092 + 0.154 = 0.246$ is near to the sum of
atomic radii $0.172 + 0.073 = 0.245.$ This implies that the Na atoms
can demonstrate "atomic" properties even in the NaCl. The
mentioned above ionic radii are rather arbitrary values \cite{bib7}. The
ionic radius values, calculated later quantum-mechanically \cite{bib8},
appeared to be noticeably less than the previously assumed: $0.087
+ 0.074 = 0.151 < 0.245.$ This confirms reality of the atomic
states even in such stoichiometric compound as NaCl (see, for
example, the paper by Laves in \cite{bib6}). This is all the more valid
for non-stoichiometric compounds. This approach to Na-WO$_{3}$ changes
the electronic states picture in the crystal used in \cite{bib4}.

Stretching of the monovalent metal lattice by the matrix can lead
to forming of the specific unique state intermediate between metal
and dielectric \cite{bib5}.
\paragraph*{\rm{II.}}
\label{pII}
        If the lattice of monovalent metal atoms, located in the voids of diameter $D$ larger than the atomic diameter $d_{0}$  $(D - d_{0} = h > 0)$, is stretched, then the atoms can occupy one of a few positions. Doubling of the metal sublattice period can occur in this case due to forming of the Me$_{2}$ - molecules from the atoms in the neighboring voids. Their binding energy can be essentially lower than the metal cohesive energy. Nevertheless, the symmetry lowers and the energy gap appears in the vicinity of the Fermi level. The question arises if this gap is just dielectric or it is related to superconductivity \cite{bib9}.
\subparagraph{1.}
\label{pII.1}
    Monovalent metals are not usually superconductors. However, superconductivity of the Na-WO$_{3}$ compound appears due to introducing of the Na monovalent metal atoms into the WO$_{3}$ matrix. This controversy can be eliminated, if we take into consideration that the period doubling is accompanied by forming of the molecular-like Na$_{2}$ structures; their binding energy is determined by a contribution of the molecular orbital with two paired electrons.

We can calculate the energy gain at transition from usual
monovalent metal to the "slightly stretched" one and estimate the
energy gap magnitude. The cohesive energy of the stretched metal
Na is $E_{s} \sim 1 /D$, while we have for the usual Na metal $E_{0}  \sim 1/d_{0}$.
The cohesive energy difference at the displacive phase transition
Na$_{1}$- Na$_{2}$ is about
\begin{equation}
\label{f1}
\Delta E = E_{0}h/(D-h) \sim E_{0}h/d_{0}; \quad  ( D-d_{0} ) = h > 0
\end{equation}
Using the conventional values for parameters of the system {Na-WO$_{3}
:E_{0} = 1$\rm{\,eV},}\; {$D = 0.378$\rm{\,nm},}\; {$d_{0} = 0.372$ \rm{\,nm},}\; {$h = 0.006$\rm{\,nm}}, we obtain
$\Delta E = 0.016 \rm{\,eV} \sim 187$K. The molecular binding energy 0.35\,eV is lowered up
to 0.016\rm\,eV when Na$_{2}$ molecule embedded into the WO$_{3}$ \rm medium. This well corresponds to dielectric susceptibility of about 20 both for NH$_{3}$ and \mbox{NdBa$_{2}$Cu$_{3}$O$_{7}$}

   The available experimental data on Na-WO$_{3}$ are in a good accordance with the considered above scenario. For example, the symmetry lowering by the lattice period doubling is really observed at transition from WO$_{3}$ to Na-WO$_{3}$ according to the x-ray study \cite{bib1,bib4}. Tetragonal phase domains appear in the cubic WO$_{3}$ \cite{bib1,bib4}. The superconducting energy gap has been measured by the electron spin resonance technique and appeared to be equal 160-180K \cite{bib4}. This confirms the presented above estimate for $\Delta E,$ which has been done using only the crystallographic parameters of the system and the energy $E_{0}.$ In addition, a suppression of the superconductivity at $h < 0$ (see eq.\ref{f1}) has been found experimentally as well. The compounds Rb-WO$_{3}$ and Cs-WO$_{3}$ demonstrate superconductivity only below ~ 6K \cite{bib4}. Rb ($d_{0} = 0.496$\,nm) and Cs ($d_{0} = 0.532$\,nm) atoms are strongly compressed in the WO$_{3}$ voids. Cesium!
, for instance, have $T_{c} \sim 2$K at $P = 10$ GPa \cite{bib10}. The parameter $h > 0.06$\,nm for the Li or Ag atoms ($d_{0} \sim 0.308\,\rm{nm,}$  $\Delta E / E_{0} \sim 0.1$), but it is related to surprisingly high $T_{c}$. At a large crystallographic misfit, phase transitions of a different kind are possible, which change the matrix structure.
\subparagraph{2.}
\label{pII.2}
    Repeated attempts to discover superconductivity of the Na-NH$_{3}$ solutions have been made according to \cite{bib11}. The dramatic story of this study is described in \cite{bib12}. Such solutions can be metastable at various stages of the phase segregation and contain groups Na$_{2}$, Na$_{3}$, ..... \cite{bib5}. Possibly, just this feature allowed to freeze-in the magnetic flux in the sample only using fast cooling. Numerous attempts to obtain "diluted" metals vaporizing them at the NG atmosphere failed because of the fast phase segregation at the substrate.
 \subparagraph{3.}
 \label{pII.3}
   A similar scenario of the Jahn-Teller-Peierls transformations
can be effected also in the case, when the introduced component
stretches the matrix (see point I. 1) and its lattice constant
$D_{0}$ increases up to $D$. We have in this case
\begin{equation}
\label{f2}
\Delta E \sim E_{0}h / D_{0} ; \quad  ( D-D_{0} ) = h  > 0
\end{equation}
The fullerides Me$_{n}\rm{C}_{60}$ give an example of such system. Clusters
Me$_{n}$\rm{C}$_{60}$, located in the octahedral voids of the molecular crystal C$_{60}$
expand the lattice of C$_{60}$ spherical molecules and the lattice
constant increases from $D_{0} =1.37$\,nm to $D$ due to a decrease of the
packing density \cite{bib2}. The relation (\ref{f2}) is valid for a wide variety
of the alkali metals for $n = 3$.  $T_{c} = 40$K (Fig.3 of Ref.\cite{bib2}) at $D = 1.47 $\,nm,  $E_{0} \sim 0.1$ eV.  $(E_{0}$ - part of intermolecular interaction
ehergy determined by the electronic doping).  Moreover, if $n = 4$ or 6, the  Me$_{n}$C$_{60}$ crystal is dielectric \cite{bib2}. The transition temperature $T_{c}$ of fullerides is
proportional to $h/D_{0}$. ( For the tungsten-bronzes with Na, Rb, or Cs,
$T_{c} > 0$ at $h > 0,$  and $T_{c} \sim 0$ for $h < 0 )$.
\subparagraph{4.}
\label{pII.4}
  The high temperature superconductors of the YBCO type also have a perovskite matrix composed from the CuO$_{5}$ and CuO$_{6}$ polyhedrons with voids filled by the metal atoms. For instance, $D_{a} - d_{0} = 0.39 - 0.36 = 0.03 \rm{\,nm;} \; D_{b} - d_{0} = 0.02$\rm{\,nm for Nd atoms in} \mbox{NdBa$_{2}$Cu$_{3}$O$_{7} \: (T_{c}\sim 95$K)}. The Cooper pair binding energy can also depend upon dynamics of the "stretched" atoms in the matrix voids.

The totality of considered experimental data on superconductivity
of a variety of systems of the matrix-filling type does not
contradict a scenario, where a weak modulation of the monovalent
metal lattice constant ($h/d_{0} \sim \Delta E/E_{0} \sim 0.016$ for Na-WO$_{3}$) leads to forming
of the bound electronic states of the molecular type and the
energy gap. There are parallels between this result and both weak
coupling BCS superconductivity and the Peierls transition, which
is usually assumed to be only a transition to the dielectric state .

\end{document}